\documentclass[pra,nofootinbib]{revtex4}

\usepackage[titletoc,title]{appendix}
\usepackage[utf8]{inputenc}
\usepackage[T1]{fontenc}
\usepackage[english]{babel} 
\usepackage{amsmath}
\usepackage{amsfonts}
\usepackage{amssymb}
\usepackage{amsthm}
\usepackage{euscript}
\usepackage{tikz}
\usepackage{url}
\usepackage{algorithm,algorithmic}   

\newtheorem{construction}{Construction}

\begin{document}

\title{Measurement based fault tolerance beyond foliation}

\makeatletter
\let\@fnsymbol\@arabic
\makeatother

\author{Naomi Nickerson}
\affiliation{PsiQuantum}

\author{H\'ector Bomb\'in}
\affiliation{PsiQuantum}

\begin{abstract}
In order to build a scalable quantum computer error correction will be required to reduce the impact of errors. Implementing error correction in the framework of measurement based computation manifests itself as the construction of fault tolerant cluster states (FTCSs). While any 2-dimensional stabilizer code can be used to construct a FTCS through the process of foliation, here we find FTCSs that cannot be constructed through foliation of a stabilizer code, and identify new examples of self-dual codes that can still be implemented in a 2-dimensional physical architecture.

\end{abstract}

\maketitle

\section{Introduction}

Any design for a scalable quantum computer must include quantum error correction to reduce the impact of errors. Topological codes are the most viable candidates currently known due to their high thresholds and geometric locality. Amongst these, Kitaev's surface code~\cite{kitaev1997quantum,kitaev2003fault,dennis2002topological} is by far the most commonly studied in both theory and experiment.

\begin{figure}[h!]
\includegraphics[width=0.8\columnwidth]{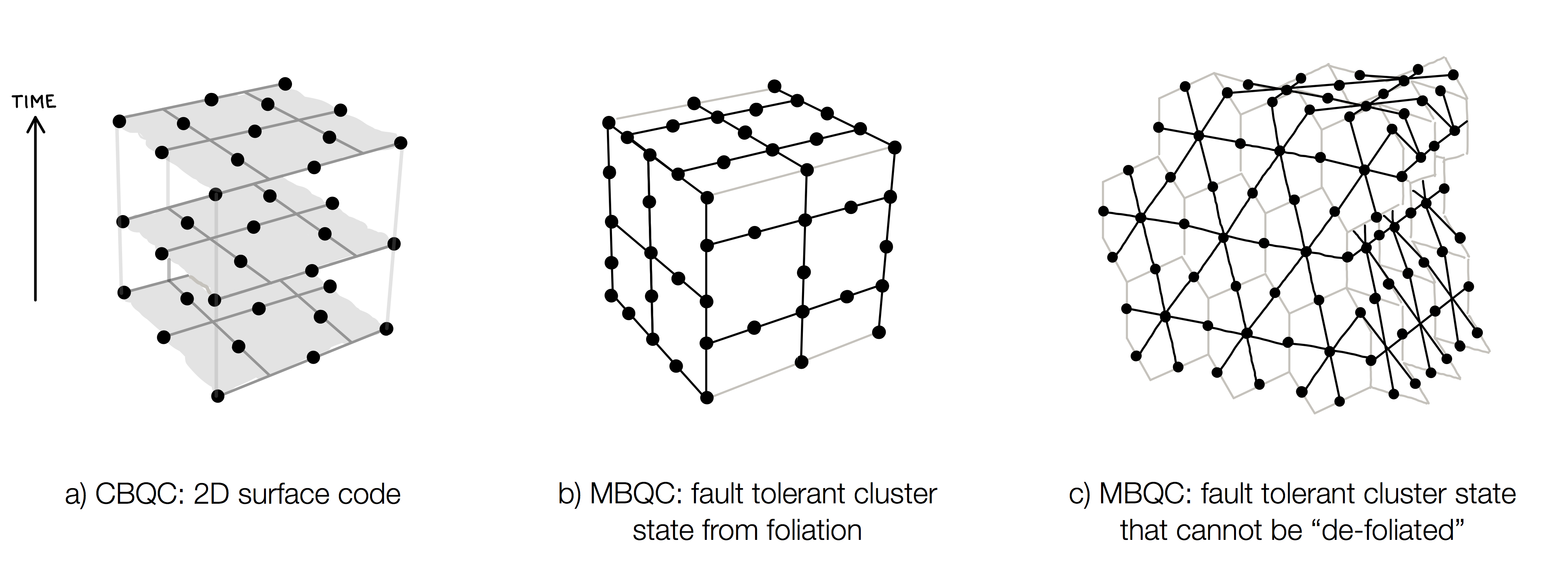}
\caption{{\bf Fault tolerant cluster states.} a) A representation of the operation of the square surface code in space + time. b) The fault tolerant cluster state resulting from foliation of the square surface code in a). c) One example of a fault tolerant cluster state that we introduce that cannot be `de-foliated'. This example is based on the geometry of the diamond lattice. This state cannot be represented as a single 2-d code undergoing multiple rounds of stabilizer measurement.}
\end{figure}

The surface code can be implemented both in circuit based quantum computation (CBQC), and measurement based quantum computation (MBQC), which provide two different scenarios for the physical primitives allowed for processing quantum information. Under CBQC qubits can be prepared, measured, and acted upon with 2-qubit gates. In this scenario the surface code has a 2-dimensional structure, on which parity-check measurements are made of small groups of nearby qubits in order to detect errors. Since measurements on the state will also be error-prone, multiple rounds of measurement must be made in order to successfully achieve fault-tolerance. In MBQC, a large entangled resource is prepared, following which all interaction with the state is through single qubit measurements~\cite{briegel2009measurement}. The MBQC implementation of the surface code is the 3-dimensional cluster state originally introduced by Raussendorf, Briegel and Harrington~\cite{raussendorf2005long,raussendorf2006fault,raussendorf2007fault}, which we refer to as the cubic fault tolerant cluster state (FTCS). More generally any 2-dimensional CSS code can be converted into a 3-dimensional cluster state through the process of foliation~\cite{foliation}. Physically the measurement based approach can be implemented both in matter based qubits~\cite{paler2014mapping}, or in systems where there is no direct access to two-qubit gates, such as linear optics~\cite{kieling2007percolation,gimeno2015three,li2015resource}. 

In both approaches it is the 3-d array of syndrome information obtained via measurement that is necessary to achieve fault tolerance. In CBQC time acts as the third dimension, whereas in MBQC the state itself is 3-d. There is a direct correspondence between the two pictures but, under this correspondence, the MBQC picture is naturally more flexible: it allows us to abandon the notion of a fixed set of physical qubits encoding logical information. Rather the cluster state, along with suitable measurement patterns, represents a fault-tolerant map between two codes. MBQC gives us a clear insight into the freedom in all three dimensions. This freedom allows the construction of fault tolerant cluster states that cannot be considered as foliated 2-d stabilizer codes. In particular, moving beyond foliated codes presents us with the option of decoupling the geometry of the X- and Z- parts of the code that for any 2-d surface code are constrained to be dual to one another. To find examples of these states we introduce a method, which we call {\em splitting}, through which an existing FTCS can be modified to create a new one with different properties. 

Part of the original motivation for exploring different cluster state geometries was to identify states that could be more tolerant to loss and errors, in particular for photonic quantum computing where there are no particular geometric constraints on the state since non-local connectivity is readily available. In all the examples we introduce here the thresholds for erasure and Pauli error under a phenomenological error model are indeed significantly higher than those of any foliated surface code (Table~\ref{table:summary}). We must be careful in interpreting these numbers, since the phenomenological error model does not account for the difficulty of state preparation. A cluster state with a higher bond-degree (valence) will often require more operations to prepare, and consequently accumulate more noise. We observe a trade-off between the cluster state valence and the error tolerance of the states. Such a trade-off can only be fully analysed with respect to a specific physical architecture and a detailed error model, but we can compare the states more fairly by applying a weighted phenomenological error model, which accounts for the valence of the cluster state. The Pauli error thresholds under this model are shown in the last column of the table.

\begin{table}[]
\centering
\begin{tabular}{p{5cm}|p{2cm}|p{2.2cm}|p{2.5cm}|p{3cm}}
   & Cluster state bond degree & Erasure Threshold & Phenom. Pauli Threshold &  Weighted Phenom. Pauli Threshold\\ \hline
Cubic FTCS (foliated surface code)& 4 & 24.9\%   & 2.6\%   & 0.55\%      \\ \hline
Diamond FTCS& 6  & 39\%    & 5.2\%   & 0.87\%      \\ \hline
Doubled-edge cubic  FTCS& 8  & 50\%     & 7.5\%   & 0.94\%      \\ \hline
Triamond FTCS& 10 & 55\%    & 9.5\%   & 0.95\%   
\end{tabular}
\caption{\label{table:summary}Thresholds against erasure and Pauli error are shown for the cubic FTCS~\cite{UF,barrett2010fault} (foliated surface code), and the three new cluster state constructions we introduce here. Estimates of Pauli thresholds are determined using a simple union-find decoding algorithm~\cite{UF} with a phenomenological error model. The erasure thresholds reported are optimal, but we note that the Pauli thresholds are lower bounds since we do not use an optimal decoder. For the cubic FTCS the optimal performance is known to be 3.3\%~\cite{ohno2004phase}, and 2.9\% can be achieved an efficient max-likelihood decoder~\cite{wang2003confinement}. The final column shows the Pauli threshold under a weighted phenomenological error model, where each lattice site suffers a Pauli error with probability $p z$, where $p$ is a unit of error rate, and $z$ is the valence of the site in the cluster state.  
}
\end{table}{}

\medskip

We begin in the next section by introducing the cubic FTCS, and how it can be generalized to a larger class of fault tolerant cluster states. We then introduce the method of splitting, and how it can be used to produce new examples of states that are also self dual, and characterize their properties.

\section{The cubic FTCS}
\label{sec:raussendorf}
We begin by briefly reviewing the cubic FTCS~\cite{raussendorf2007fault} (foliated surface code). The state is a 3-d cluster state lattice construction, which, when measured in a suitable way, allows for measurement-based fault tolerant quantum computation. In the bulk of the lattice, all the qubits are measured in the X-basis, the measurement outcomes can be combined to construct syndrome information which can be decoded to correct for errors. The corrected measurement outcomes can then be used to reconstruct logical correlation operators of the code. For computation, the measurements in the bulk must be combined with suitable boundaries for the code. Here we focus only on the fault-tolerant property of the bulk lattice, and do not discuss the boundaries. How computation can be achieved given a fault tolerant cluster state is described in detail in refs.~\cite{raussendorf2007fault, raussendorf2007topological}, and these methods apply directly to our new constructions. 

The cubic FTCS is a large and complex quantum state, but it has topological properties that allow it to be concisely represented geometrically. There are two distinct graphical representations of the state that we will use here. The first is the representation of the cluster state as a graph to represent the structure of the entanglement. The second is a representation of the geometry of the stabilizers as a cell complex. Our aim in this section is to describe the cubic FTCS and its geometric properties, which will then allow us to use only the geometric representation in order to explore new constructions.

\subsection{Description of the state}
Graph states are a class of quantum state that can be defined by a set of edges and vertices, $G_C = \{E_C,V_C\}$. Each vertex of $G_C$ represents a qubit, and each edge represents an entangling bond between the two qubits it connects. The graph state is a stabilizer state defined by stabilizer generators associated with each vertex of the graph, $i \in V_C$,

$$S_i = X_i \bigotimes_{j\in nb(j)} Z_j.$$

\begin{figure}
\includegraphics[width=0.75\columnwidth]{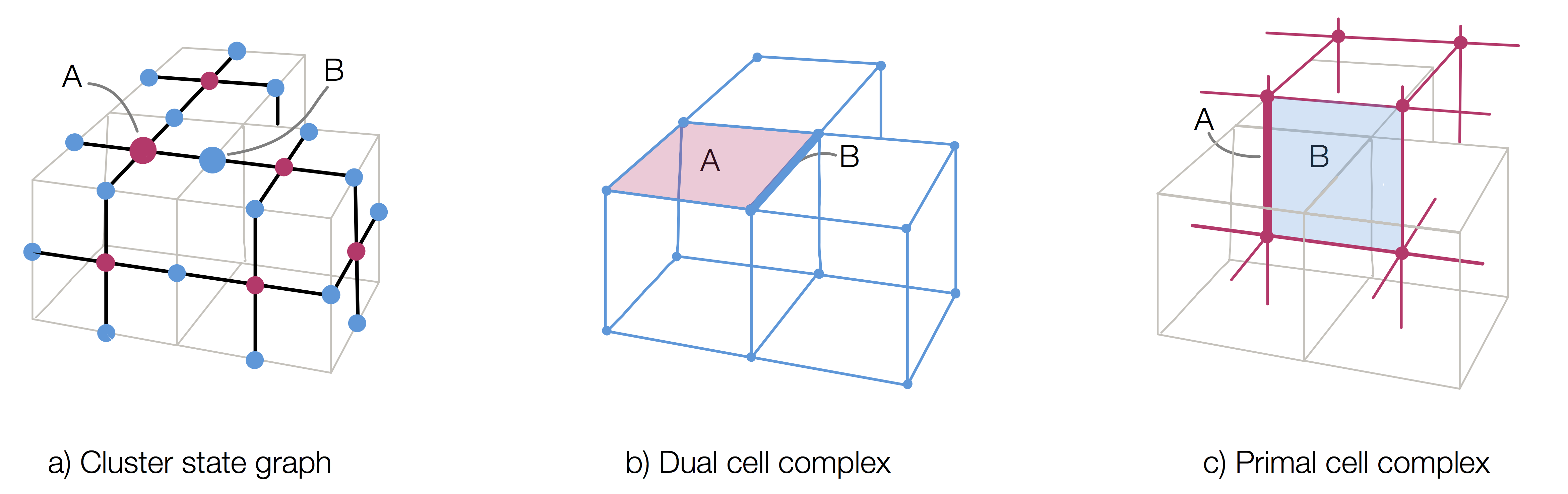}
\caption{~\label{fig:lattice_duality} Geometric representations of the cubic FTCS. Two qubits, A and B, are highlighted to show the relationship between the different representations. A is a primal qubit in the cluster state, a face in the dual lattice, and an edge in the primal lattice. B is a dual qubit in the cluster state, an edge in the dual lattice and a face in the primal lattice. a) Three unit cells of the graph state. Primal qubits are colored pink, and dual qubits are colored blue. Black lines indicate the edges of the graph state. b) The dual cell complex, faces correspond to primal qubits, and edges correspond to dual qubits. c) The primal cell complex. Edges correspond to primal qubits and faces correspond to dual qubits. 
}
\end{figure}

Where $nb(j)$ is the neighborhood of node $j$, that is, all nodes that are connected directly to node $i$. The terms {\em graph state} and {\em cluster state} are sometimes used synonymously. Here we refer to any state that can be described as above as a {\em graph state}, and reserve {\em cluster state} for larger graph states that have fault-tolerant properties. The cubic FTCS is a cluster state, and several unit cells of the lattice are shown in Figure~\ref{fig:lattice_duality}a). Every qubit of the state is identical, in that they each have four bonds and an identical neighborhood. The lattice is bipartite, and we call the two distinct sets of qubits {\em primal} and {\em dual} qubits. The stabilizers of the state can be divided into those which are centered on a primal qubit $S^P$ and those with are centered on a a dual qubit $S^D$. The stabilizers $S^P$ and $S^D$ form subgroups of $S$. In Figure~\ref{fig:lattice_duality}a) the primal qubits are colored pink, and the dual qubits are colored blue. 

The cluster state is defined on a cubic cell complex as shown in Figure~\ref{fig:lattice_duality}, defined by a set of cells, $C$, faces, $F$, edges, $E$ and vertices, $V$. The cell complex has a corresponding dual, $\{\bar{C},\bar{F},\bar{E},\bar{V}\}$, where $\bar{C}=V$, $\bar{F}=E$ and so on. We choose to name the two cell complexes by the duality class of the qubits that lie on their edges as this will be more convenient terminology when we come to error correction. Figure~\ref{fig:lattice_duality}b) shows the dual cell complex where the edges represent dual qubits and the faces represent primal qubits. In the primal cell complex which is shown in~\ref{fig:lattice_duality} dual qubits correspond to faces, and primal qubits correspond to edges.  The coloring in Figure~\ref{fig:lattice_duality} shows the relationship between the three representations, with two specific qubits highlighted. 

The dual (or primal) cell complex provides a full description of the quantum state. Given a dual cell complex the graph state can be straightforwardly constructed in the following way:
\begin{construction}
\label{cs_construction}
\item{1. Associate a dual qubit with every edge and a primal qubit with every face of the lattice}
\item{2. Connect with a graph bond the qubit at the center of each face with those around its boundary}
\end{construction}

\subsection{Geometry of the stabilizer group}

\begin{figure}
\includegraphics[width=0.7\columnwidth]{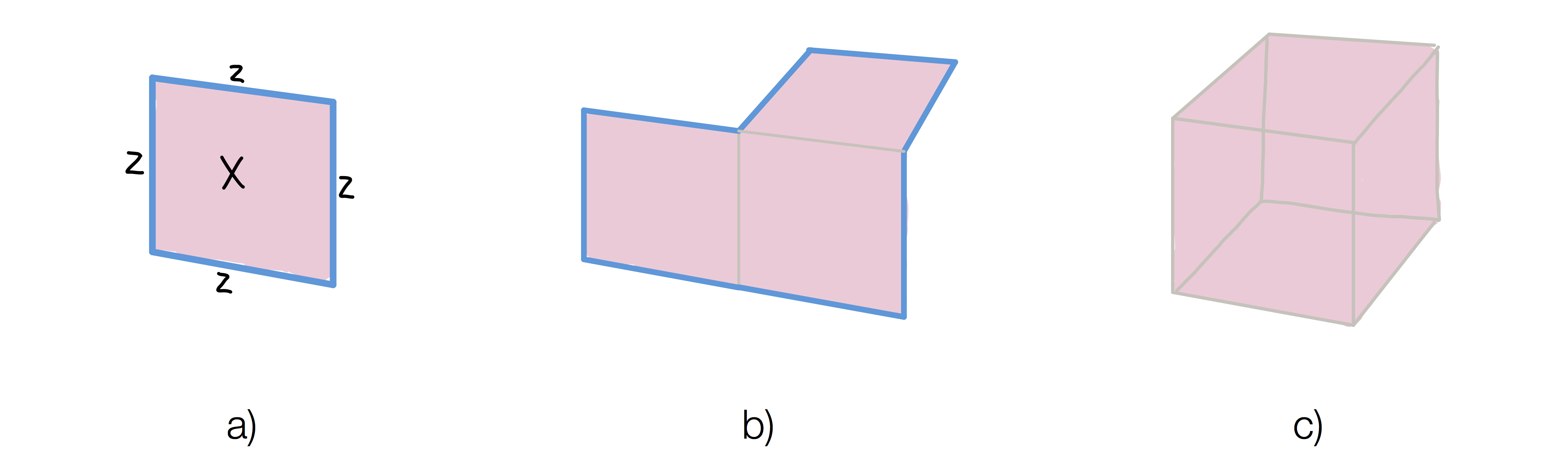}
\caption{\label{fig:stabilizer_types}. Types of stabilizers. a) A single face of the code lattice represents a single cluster-state stabilizer. The surface of the face represents the X-part of the stabilizer, while the boundary of the face represents the Z-part. b) An open stabilizer. The picture shows the stabilizer generated by combining three face stabilizers. Where two of the faces share a boundary the Z-parts of the stabilizers cancel, leaving a resulting stabilizer that has the same topology as a). The surface is made up of Pauli-X operators, while the boundary supports Pauli-Z operators. c) A closed stabilizer. The result of combining the six face stabilizers surrounding a cubic cell. All Z-components of the face stabilizers cancel leaving a closed surface supporting only Pauli-X operators. }
\end{figure}

In the primal lattice there is a dual qubit, $i$, and therefore a cluster state stabilizer, $S^D_i$ associated with each primal face, $f$. Similarly there is a primal cluster state stabilizer, $S^P$, associated with each dual face, $\bar{f}$. A useful way to think about each face, $f$, is that it represents the X part of the stabilizer while its boundary, $\partial f$, represents the Z part of the stabilizer. This is shown in Figure~\ref{fig:stabilizer_types}a). This gives us an alternative geometric description of the cluster-state stabilizers as,
\begin{align}
\label{eq:cluster_stabilizers}
S^D_f = X_f \otimes_{e \in \partial f} Z_e \\
S^P_{\bar{f}} = X_{\bar{f}} \otimes_{\bar{e} \in \partial \bar{f}} Z_{\bar{e}}.
\end{align}

If we take the product of multiple face stabilizers of the same duality class, we can identify two distinct types of stabilizers, which are indicated in Figure~\ref{fig:stabilizer_types}b) and c). In the first case, we have a product of faces which forms an {\em open surface}, in the sense that it has an X-like surface, with a Z-like boundary. In the second case we show a product of faces which forms a {\em closed surface}, in this case there is no Z-like boundary. 

In the cubic FTCS there is a closed stabilizer associated with every cell, $c$, of the primal lattice, which is formed by taking the product of all the faces of the cell. These stabilizers will provide the syndrome information that allows error correction to be performed. Similarly, for the dual lattice there is a closed stabilizer for every cell, $\bar{c} \in \bar{C}$. 
\begin{align}
\label{eq:code_stabilizers}
S^P_c = \bigotimes_{f\in \partial(c)} X_f \\
S^D_{\bar{c}} = \bigotimes_{\bar{f} \in \partial(\bar{c})} X_{\bar{f}}.
\end{align}

The closed stabilizers form subgroup of the cluster state stabilizer group, $\mathcal{S}^P_C \subset \mathcal{S}^P$. The existence of this group of closed stabilizers is what gives the cubic FTCS its fault tolerant properties. In the bulk of the lattice all qubits are measured in the X basis, which means the only stabilizers that can be reconstructed after measurement are those which contain no Z-component.

\subsection{Syndrome graph and error correction}

\begin{figure}
\includegraphics[width=0.4\columnwidth]{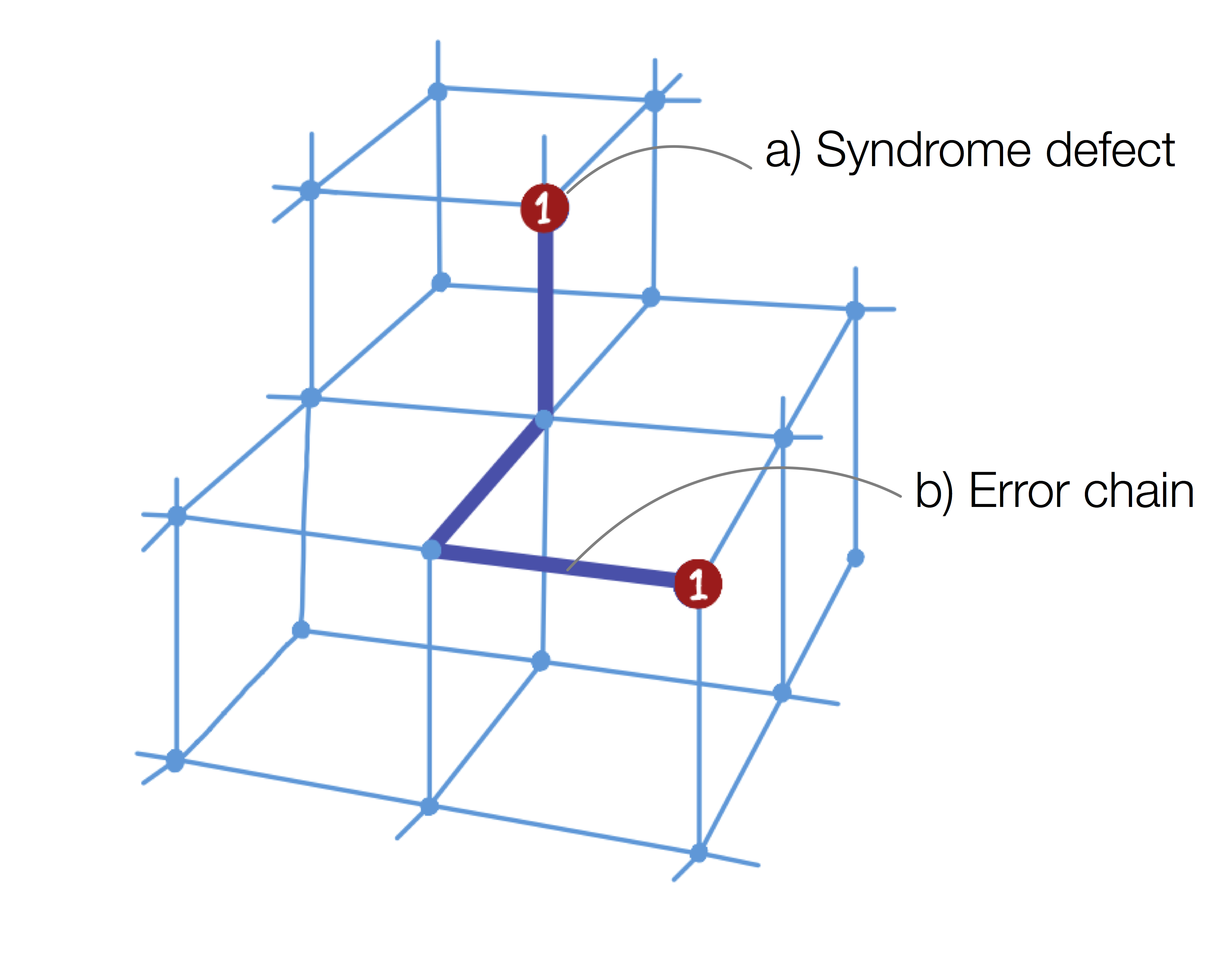}
\caption{\label{fig:syndrome_graph} Syndrome Graph and error correction. A section of the dual lattice is shown, the vertices and edges of the dual lattice define the dual syndrome graph. The edges correspond to qubits, and the vertices correspond to code stabilizers. Stabilizer outcomes are reconstructed from taking the product of single qubit measurements in the X basis of all qubits incident to a given vertex. If there are no errors on the qubits, all stabilizer outcomes are eve
n (0). If a qubit suffers a Pauli error, the value of the stabilizer associated with the vertex at each end of its corresponding edge in the syndrome graph becomes odd parity (1). A chain of errors on the syndrome graph causing an odd parity syndrome outcome at each end of the chain. The figure shows an example of an error chain of length three (dark blue edges), and the two odd parity syndrome outcomes it causes (red circles). 
}
\end{figure}

Error correction is performed by measuring all qubits (in the bulk) in the X basis, and then reconstructing the outcomes of the code stabilizers. We have so far described the code stabilizers as being associated with cells of the lattice, but in order to understand the error correction process it is much more convenient to move to the dual picture, where stabilizers are instead associated with vertices and qubits are associated with edges. The set of edges and vertices is known as the {\em syndrome graph}. An example of the dual syndrome graph is shown in Figure~\ref{fig:syndrome_graph}, it is simply the 1-skeleton of the dual cell complex.  

The outcomes of the single qubit measurements are associated with edges of the graph, and the stabilizer values associated with the vertices can then be reconstructed by computing the parity of all the edges incident to the vertex. If there are no errors in the measurement outcomes then all of these reconstructed values must be even (0). But in the case of a qubit measurement error, some vertices will become odd (1). The values of all the stabilizer outcomes is known as the syndrome. An example is shown in Figure~\ref{fig:syndrome_graph} where three measurement errors have occurred in a chain, causing an odd parity stabilizer outcome at each end of the chain. 

We note that the cubic structure of the syndrome graph is identical for the cubic FTCS and fault tolerant operation of a 2-d surface code. In the FTCS the syndrome values assigned to each vertex come from combining the 6 single qubit measurement outcomes from the incident edges, whereas in the 2-d surface code the syndrome values come from combining two stabilizer measurements in consecutive time steps. Fundamentally the schemes are equivalent, they just differ in their implementation.

\subsection{Logical operators}
The logical correlation operators of a FTCS are made up of an extensive membrane of connected faces of the cell complex. As in Figure~\ref{fig:stabilizer_types}b) the surface of the membrane is defined by a set of Pauli-$X$ operators on faces of the cell complex. These operators are sometimes referred to as {\em correlation surfaces}. Suitable boundary conditions on the cluster state allow these correlation surface to terminate, or be deformed in such a way as to allow fault tolerant computation. This is described in detail in the original work of Raussendorf {\em et al.}~\cite{raussendorf2007topological,raussendorf2007fault}, and these methods have subsequently been expanded upon~\cite{herr2017lattice,brown2017poking,horsman2012surface}. The important feature relevant to our new FTCS constructions is that the logical operators are reconstructed from the same single qubit measurements that allow fault tolerance, no other operations are required in the bulk of the lattice in order to compute. 

\subsection{Summary}
To summarize, we can define the cubic FTCS as a cubic cell complex made up of cells, faces, edge{}s and vertices: $\{C,F,E,V\}$. Each element of this cell complex and its dual correspond to features of the code, as detailed in Table~\ref{table:cell_complex}. We can construct the cluster state directly from this cell complex. 

\begin{table}[h]
\centering{}
\begin{tabular}{l|l|l}

                       & Primal lattice & Dual lattice \\ \hline
Primal closed stabilizers & Vertices       & Cells        \\ \hline
Primal qubits          & Edges          & Faces        \\ \hline
Dual qubits            & Faces          & Edges        \\ \hline
Dual closed stabilizers  & Cells          & Vertices     \\ \hline
\end{tabular}
\caption{\label{table:cell_complex}Elements of the cell complex defining a FTCS.}
\end{table}

\section{Fault tolerant cluster states} 

All the properties of the cubic FTCS that we described in the previous section can be immediately applied to any cell complex, regardless of its geometry. Let us consider constructing a cluster state according to Construction~\ref{cs_construction} on an arbitrary 3-d cell complex $\mathcal{C}=\{C,F,E,V\}$, that completely fills space\footnote{in geometry such a cell complex is known as a honeycomb}. This cluster state has the following properties:

\begin{itemize}
\item{\bf Cluster state stabilizers}: There is a primal (dual) cluster state stabilizer associated with every face of the dual (primal) cell complex according to Eq.~\ref{eq:cluster_stabilizers}.
\item{{\bf Code stabilizers:} There are primal (dual) closed stabilizers associated with every cell of $\mathcal{C}$, according to Eq~\ref{eq:code_stabilizers} }
\item{{\bf Duality:} A cell complex, $\mathcal{C}$ always has a dual cell complex $\mathcal{C}^*$, so if the primal lattice forms a valid primal code, then the dual lattice must also do so. This means there are both primal and dual code stabilizers. There is no requirement that the cell complex be self-dual.}
\item{{\bf Logical correlation operators:} As long as the cell complex fills space then it is always possible to find a sheet of connected faces that extends through the lattice. With suitable boundary conditions that allow the sheet to terminate, this gives us a logical correlation operator.   }
\end{itemize}

\begin{figure}
\includegraphics[width=0.6\columnwidth]{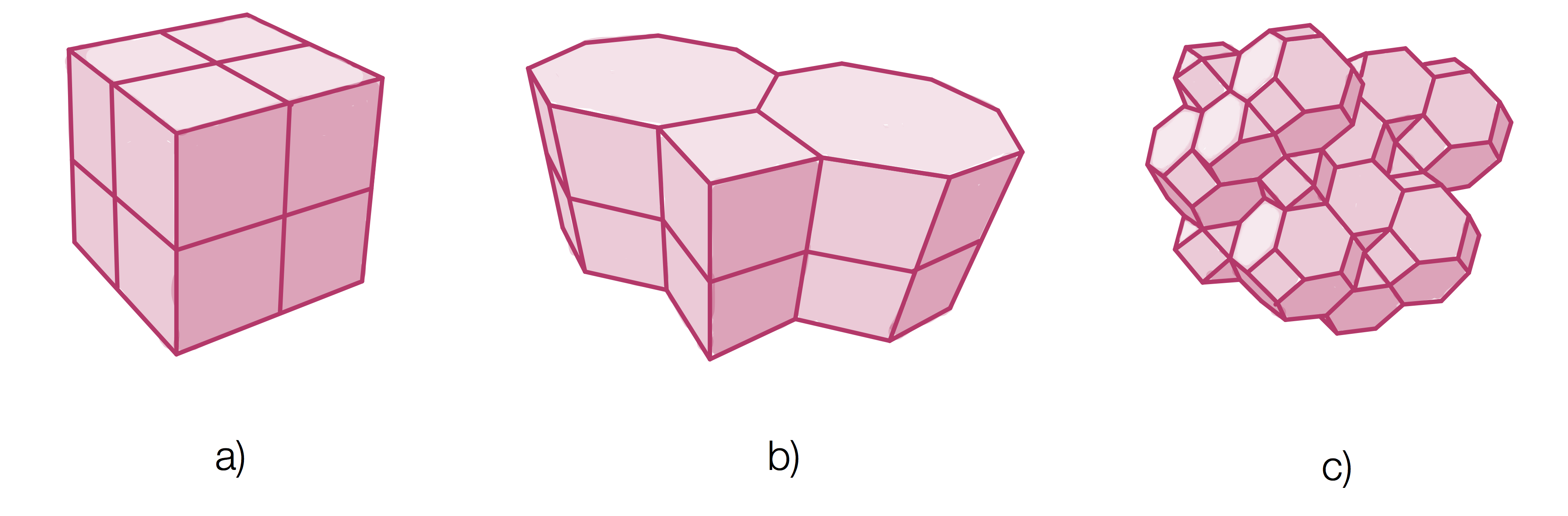}
\caption{\label{fig:honeycombs}Examples of surface-code like FTCSs. a) A cubic cell complex which corresponds to the cubic FTCS. b) A cell complex of repeated identical layers where all cells are prisms. This corresponds to the foliation of a surface code. c) An example of a cell complex without a layered structure, and therefore with no natural interpretation as a foliated code. Any cell complex that completely fills space defines a FTCS.}
\end{figure}

Several examples of cell complexes representing surface-code FTCSs are shown in Figure~\ref{fig:honeycombs}. These cluster states represent a fault tolerant map between two 2-d surface codes on the boundaries of the state. 

We comment briefly on the relationship of these states to foliated surface codes. The process of foliation~\cite{foliation} can be thought of as encoding multiple rounds of stabilizer measurement on a code into a cluster state. Foliation of the surface code on a square lattice results in the cubic FTCS, but the process can be applied to an arbitrary surface code. Any foliation of a 2-d surface code will produce an FTCS which is made up of layers of prismatic cells. An example is shown in Figure~\ref{fig:honeycombs}b). Faces of the cell complex in the $x-y$ plane may have any shape, but all faces in the $z$-direction are rectangular. In foliated codes there is an intuitive notion of an input state (the first layer of the cluster state), and an output state (the final layer of the cluster state). Both the input and output are the same 2-d surface code, and the cluster state between the two layers forms the fault-tolerant channel between them.

\section{Exploring new geometries}
Now we have identified that we can consider any cell complex as a fault tolerant code, the important question to ask is whether we can gain any advantage by deviating from the purely cubic geometry of the cubic FTCS? A major motivation for exploring states that cannot be interpreted as 2-d stabilizer codes is that we can hope to remove some of the inconvenient constraints that arise in 2-d. A lattice in 2-d completely defines its dual, so for the 2-d surface code choosing the geometric structure of the X-part of the code completely defines the Z-part. Furthermore, it is always the case that boosting the threshold of one part causes a reduction in the threshold of the other. Since the overall performance is determined by the worst of the two parts this means the optimal code (assuming uniform errors) is one that is self-dual. There is only one self-dual lattice in 2-d, the square lattice, which defines the most well known variant of the surface code. 

However, for a 3-d fault tolerant cluster state, these constraints no longer apply. For the 3-d cell complex that defines such states the definition of a primal lattice {\em does not} completely define the dual, and this gives us the powerful flexibility to vary the threshold of the X- and Z-parts independently. Essentially, it is possible to create a cluster state that forms a fault tolerant map between copies of 2-d surface codes where the lattice-geometry of the X and Z- parts of the code are not dual to each other. The case of being limited by the lower of the two thresholds still applies in 3-d, but remarkably we find multiple examples that are {\em self-dual} meaning it is possible to increase the threshold in {\em both bases}. 

Our goal is to identify a structure with a higher tolerance to errors, in order to do this it is useful to have a simple design principle to follow. Computing the threshold of an arbitrary code against Pauli error is fairly involved, and requires lengthy Monte Carlo simulations with a particular decoding algorithm. Understanding the performance under erasure errors on the other hand is very straightforward, and we believe that it serves as a good proxy for Pauli error tolerance. So we focus on how to construct a code with a high erasure tolerance. 
\begin{figure}
\includegraphics[width = 0.5\columnwidth]{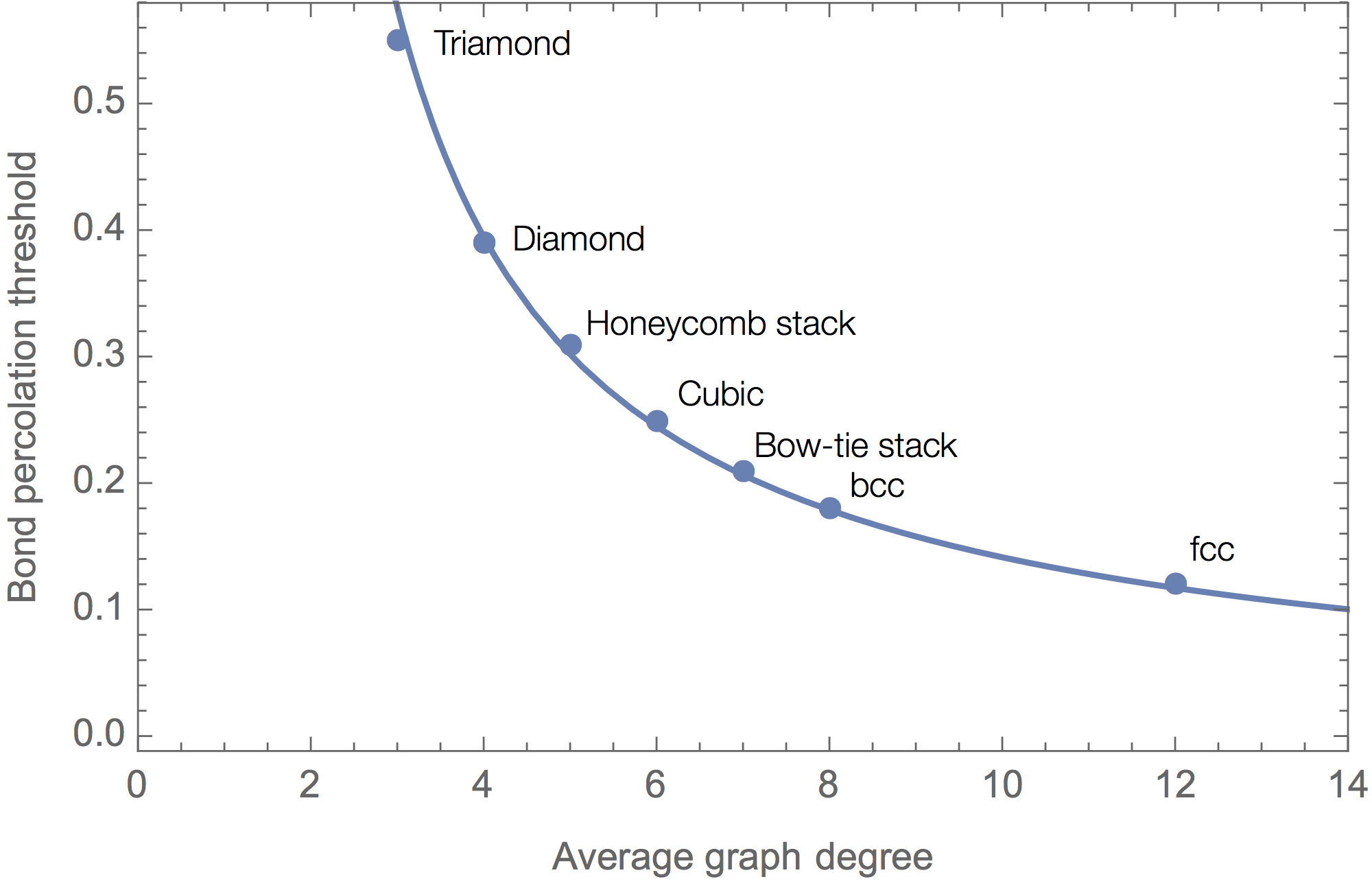}
\caption{\label{fig:percolation} Percolation Thresholds. Data points indicate some known bond percolation thresholds for regular lattices~\cite{van1997percolation,bondpercolation,lorenz1998precise,percolation3valent} plotted against the average degree of the lattice, $\bar{z}$. The solid line shows the heuristic formula, $p_{c} \sim \frac{1}{z-1}$~\cite{galam1996universal} which approximates their behavior very well. 
}
\end{figure}

The erasure threshold of a surface code, or surface-code FTCS is given simply by the bond percolation threshold of the syndrome graph. These values can be computed quickly~\cite{delfosseErasure}, and if the lattice is a standard geometry they can simply be looked up in the literature~\cite{van1997percolation,bondpercolation,lorenz1998precise,percolation3valent}. Furthermore for any 3-d lattice there is known to be a very close relationship between the degree (or valence) of the vertices of the graph and the percolation threshold. Figure~\ref{fig:percolation} plots data points of some known bond percolation thresholds in 3-d, against a heuristic fit proposed in ref.~\cite{galam1996universal}, that the percolation threshold, $p_{c} \sim \frac{1}{z-1}$~\cite{galam1996universal}, where $z$ is the degree of the vertices of the graph. This relationship it not exact, but in practice holds very closely for regular lattices. This tells us that we should be aiming to find a structure with a syndrome graph that has the lowest possible degree. We can also see an intuition about how this feature might lead to a higher Pauli error threshold, which is that a non-trivial syndrome vertex in the syndrome graph tells us that one out of its $z$ neighboring qubits has suffered a Pauli error. The smaller $z$ the more information we have about the location of the error.

Given the goal of minimizing the bond-degree of the syndrome graph we can see the potential advantage of moving away from foliated 2-d codes. For a connected graph whose vertices have uniform degree, 3 is the minimum degree in any dimension. Therefore the syndrome graph of any foliated surface code must have degree at least 5. It is only possible to reduce the degree of the syndrome graph further by moving away from a `layer-by-layer' approach.

One more important consideration when exploring new geometries is the cost of preparing the cluster state. Quantifying this cost is very difficult without reference to a specific architecture since there are many possible methods for state preparation. But the valence of the cluster state can give us a rough measure of how hard it is to build. The higher the valence, the more entangling operations each qubit must undergo during state preparation. In our cell complex picture the valence of the cluster state corresponds to the number of edges of the faces in the primal and dual lattice. So ideally we would like to identify FTCSs that have high error tolerance, and are represented by a cell complex made up only of faces with a small number of edges.

\subsection{Splitting}
\begin{figure}
\includegraphics[width=0.9\columnwidth]{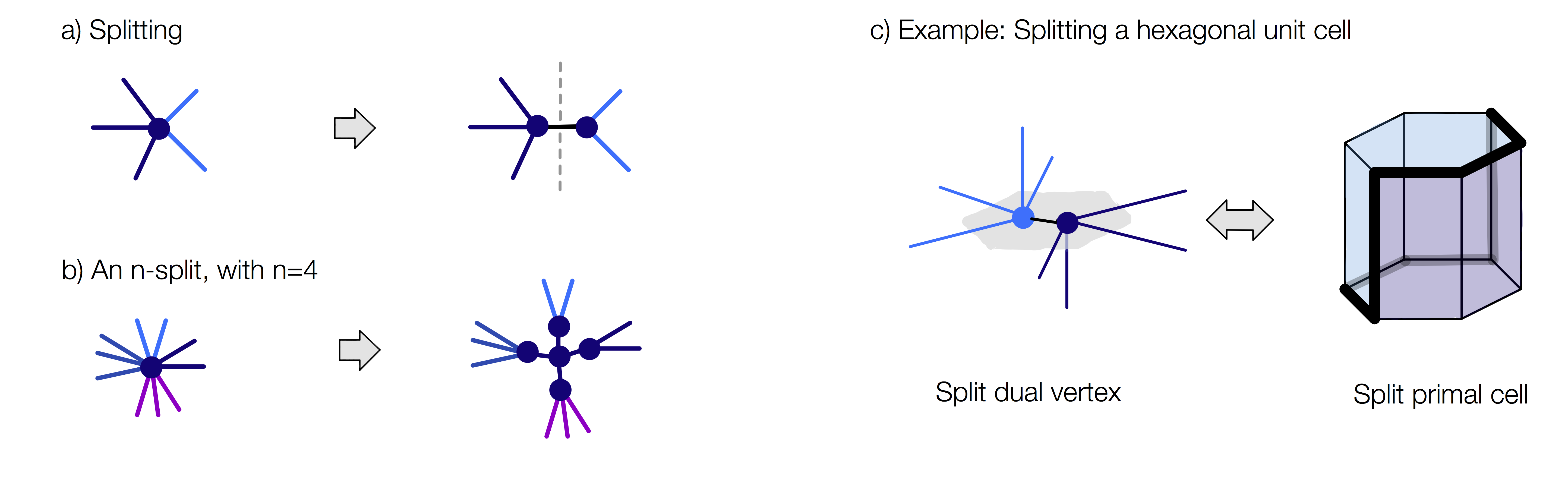}
\caption{\label{fig:splitting} Splitting the syndrome graph. a) A simple split is a bipartition of the edges incident at a vertex. The vertex is replaced by two vertices connected by a new edge. b) An $n$-split is a composition of multiple splits. $n$ new vertices and edges are added. c) The relationship between a split vertex in the dual lattice, and its representation in the primal lattice, where in this example a new face is added that bisects the cell. The outline of the new face is indicated by the bold black line.
}
\end{figure}

We now introduce a method to reduce the valence of cell complex. Let us say that we begin with some 3-d cell complex $\{C,F,E,V\}$. The primal syndrome graph is described by the 1-skeleton of this complex, $\{E,V\}$. The degree of this graph can be reduced by `splitting' vertices. The splitting process is represented in Figure~\ref{fig:splitting}a), it is defined by a bipartition of the set of edges meeting at the vertex. A split vertex is replaced by two new vertices, connected by an edge, and each new vertex maintains a subset of the original edges. The biparition of edges defining the split must be such that in the dual picture this corresponds to dividing a cell in to two connected pieces by adding one new face. 

An example is shown in Figure~\ref{fig:splitting}c) in which a dual vertex with 8 bonds is split evenly into two vertices with 5 bonds. On the primal lattice, the splitting corresponds to a new face being added that bisects the original cell. This face is not flat and its boundary is highlighted in the figure. The cell structure of the dual lattice changes, now each cell is replaced by two new cells. Importantly the primal vertices and edges are unaffected by the dual splitting.  

\medskip

To summarize, a dual split: 
\begin{itemize}
\item{Reduces the degree of the dual syndrome graph.}
\item{Changes the number of edges of dual faces.}
\item{Creates a new primal face.}
\item{Divides a primal cell into two cells. }
\item{Does not change the primal syndrome graph (1-skeleton)}
\item{Does not change the existing primal faces}
\end{itemize}

Analogously the degree of the primal syndrome graph, which is defined by the 1-skeleton of the primal cell complex, $\{{E},{V}\}$, can be reduced by splitting the primal vertices. 

\subsection{Primal and dual composition}
Primal and dual splits can be composed, because a primal split does not alter the dual 1-skeleton, and vice versa. Moreover this composition is unique, and therefore primal and dual splits commute. This means that if we apply splitting to the primal lattice, we can improve the threshold of the primal code without affecting the dual. This is the essential feature that will allow us to improve the threshold beyond that of the cubic FTCS.

\subsection{n-splits}
Splits can be composed any number of times, and the commutativity between primal and dual is a powerful tool to analyze such cases. Here we consider a particularly simple kind of composed split: n-splits. An n-split is defined by an $n$-partition of each vertex, $n \geq 3$. It introduces n new edges and n new vertices. An example of a 4-split vertex is shown in Figure~\ref{fig:splitting}b).

\section{Self-Dual Lattices}

We now apply the splitting technique to construct lattices with a low bond degree. A feature of all the lattices we introduce here is that they are not made up of convex polyhedra, that is polyhedra that can be described by a set of linear equations that define the plane of each face of the cells. In geometry such cellulations of space are not commonly studied, but for our purposes there is no reason to constrain the cells of the lattice to be convex polyhedra. 

\subsection{Example 1: Splitting the cubic graph: a degree-4 syndrome graph}

\begin{figure}
\includegraphics[width=0.9\columnwidth]{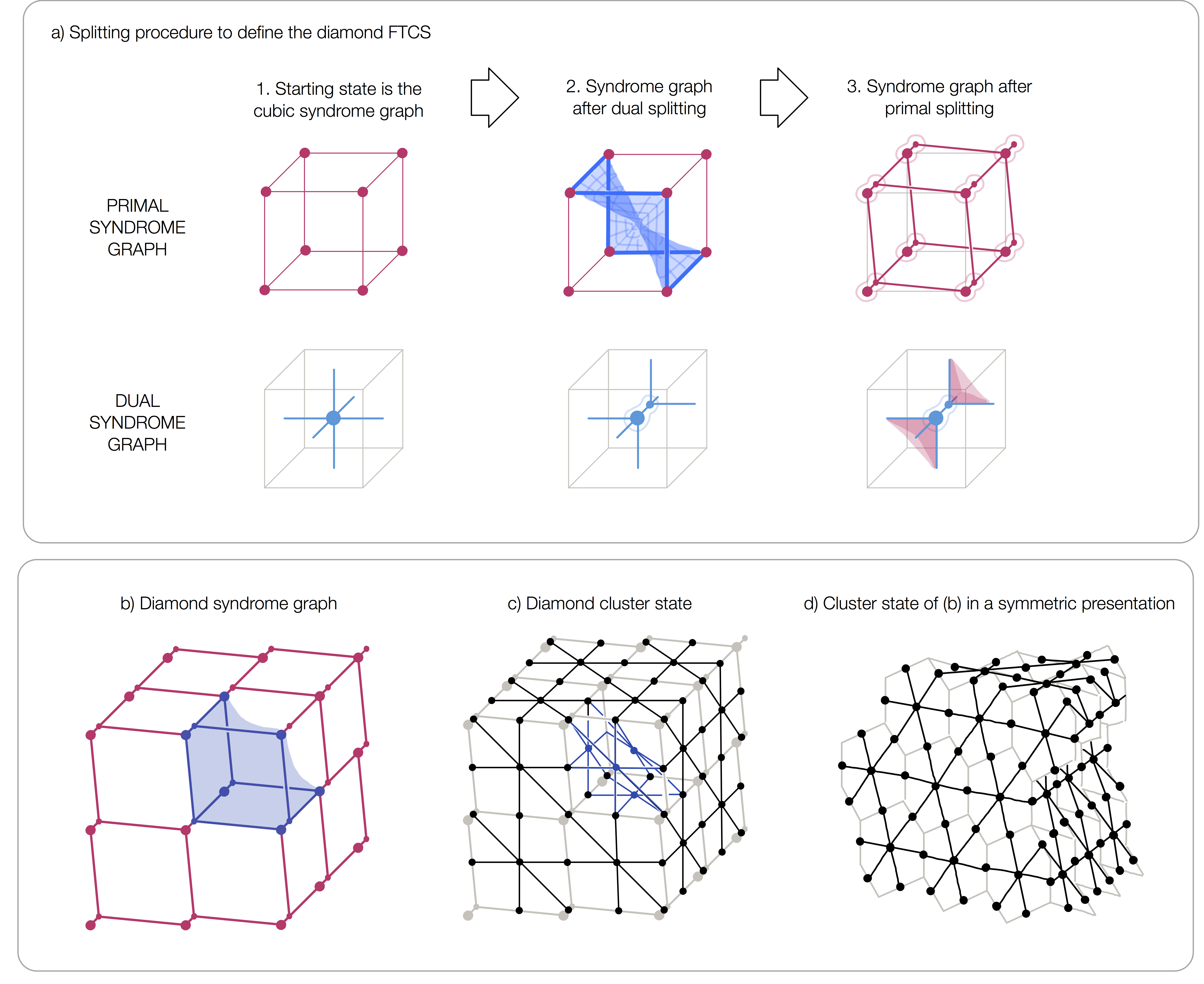}
\caption{\label{fig:diamond} The construction of the diamond fault tolerant cluster state. a) Splitting procedure to obtain a diamond-lattice syndrome graph. The primal lattice is shown in the upper row, and the corresponding dual lattice at each stage is shown in the lower row. We begin with the cubic lattice, in the first step a dual split is applied dividing each dual vertex into two 4-valent vertices. In the primal syndrome graph this corresponds to a face being added that bisects the primal cell. In the second step primal splitting divides each primal vertex into two 4-valent vertices. In the dual lattice, each new primal edge corresponds to a face.  b) The resulting structure is self-dual and is made up of identical cells that each have 4 hexagonal faces. A section of the primal syndrome graph generated from splitting is shown, along with one of the cells of the new geometry which is highlighted in blue. Each cell has four hexagonal faces, but is not a convex polyhedron. (c) The cluster state corresponding to the geometry in (b). (d) The cluster state can also be presented in a symmetric configuration corresponding to the more familiar representation of the diamond lattice. Here all bonds have equal length, and each cell is identical.
}
\end{figure}

The construction is shown in Figure~\ref{fig:diamond}. We start with the cubic cell complex. The lattice is self dual, and every vertex of the primal and dual syndrome graphs is identical and has degree 6. We first apply an even dual split such that every remaining dual vertex has degree 4. The primal 1-skeleton is unchanged, but for each cell a new face is added that bisects the cell, one such face is shown in the Figure. We then apply an identical primal split to each vertex of the primal lattice along the same axis as the dual split. 

The resulting cell complex is completely uniform and self-dual, with the 1-skeleton being the well-known diamond lattice. Each vertex is degree-4, every face is hexagonal, and every cell is formed of four hexagonal faces. This means that the cluster state that defines this code is made up of qubits which each have 6 bonds. The lattice structure and cell structure are shown in Figure~\ref{fig:diamond}. There is no configuration in which all the faces of a cell can be drawn as flat surfaces, and therefore the cells are not convex polyhedra. In Figure~\ref{fig:diamond}a) the lattice is drawn in a way that shows its relationship to the cubic lattice with the split vertices grouped together. The diamond lattice can also be drawn in the more familiar symmetric configuration where each bond has the same length, and all faces have the same shape. This is shown in Figure~\ref{fig:diamond}b) In this configuration all cells are identical.

\subsection{Example 2: 3-splitting the cubic lattice: a degree-3 syndrome graph}

\begin{figure}
\includegraphics[width=0.9\columnwidth]{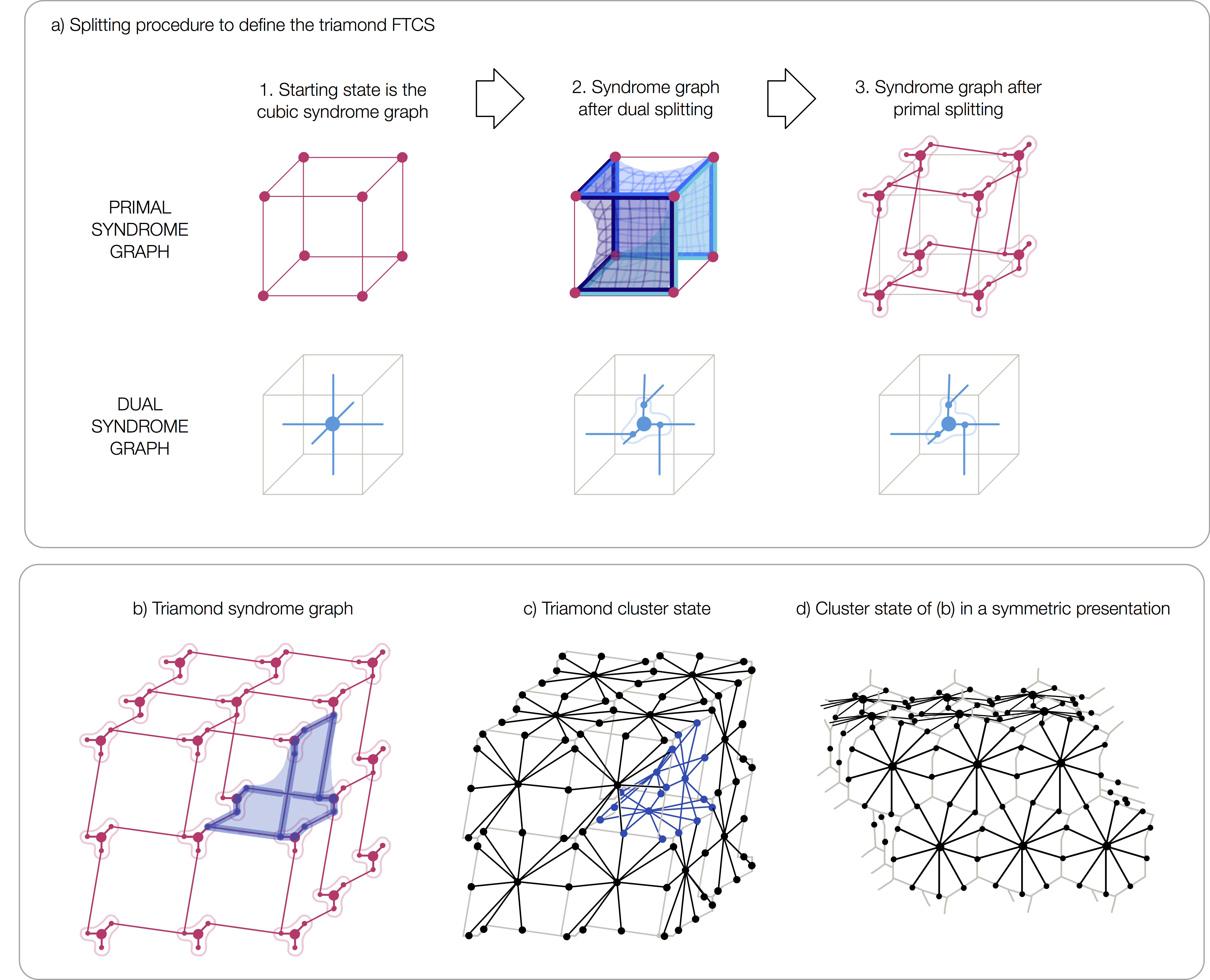}
\caption{\label{fig:triamond} Splitting procedure to obtain a triamond-lattice syndrome graph. The primal lattice is shown in the upper row, and the corresponding dual lattice at each stage is shown in the lower row. We begin with the cubic lattice (1), the primal and dual syndrome graph are shown. The dual vertex is 3-split into 4 three-valent vertices (2). In the primal picture, 3 faces are added to the cube dividing it into 4 cells, each with 3 faces. The three new faces are shown in the upper figure, and their boundaries highlighted in different colors. Finally primal splitting divides each primal vertex into four 3-valent vertices (3). In the dual lattice, each new primal edge corresponds to a face, these are omitted from the figure for clarity. b) The resulting structure is self-dual and is made up of identical cells that each have 3 decagonal faces. One such cell is highlighted in blue. c) The cluster state is shown in black, the pale grey lines indicate the geometry of the syndrome graph. d) The cluster state of (c) shown in a symmetric configuration where all edges in the syndrome graph have equal length. }
\end{figure}

We can reduce the degree of the syndrome graph further still by applying a higher-order 3-split operation to each vertex of the cubic lattice. The effect of the primal and dual splits are shown in Figure~\ref{fig:triamond}a). The dual split transforms each vertex into 4 vertices, and 3 edges, correspondingly adding three new faces to each primal cell. This is followed by an identical primal split that leads to the final lattice. The primal split also adds 3 new faces to each dual cell, but these are omitted from the figure for clarity.

The resulting structure is again self-dual, with every vertex of both the primal and dual lattices being degree-3 and every face having 10 edges. The structure of the syndrome graph is known in geometry~\cite{Sunada_crystalsthat}, and has recently been studied in several fields because of its high symmetry. It is known by several names, including the Laves graph, K4-lattice, and, as we will refer to it, the triamond lattice, because of its close relationship to the diamond structure. Tiling the unit cells yields the lattice shown in Figure~\ref{fig:triamond}b), this deformation shows its close relationship to the cubic lattice via splitting. The cluster state this new syndrome graph defines is shown in Figure~\ref{fig:triamond}c).A depiction of the lattice showing the symmetric structure, where every edge has the same length and angle of separation, is shown in Figure~\ref{fig:triamond}d). The unit cells of this structure are all identical, and made up of three decagonal faces.

\subsection{Example 3 - Uneven splitting on the cubic lattice}

The diamond and triamond FTCS both are both highly symmetric and are directly related to the cubic FTCS via splitting of every vertex. If all vertices have the same valence, then the minimum achievable is $z=3$ in order to have a connected 3-d graph. However, it is possible to further reduce the average graph degree by applying uneven splitting to add new degree-2 vertices. We now give an example of such an uneven splitting applied directly to the cubic FTCS. 

\begin{figure}
\includegraphics[width=0.9\columnwidth]{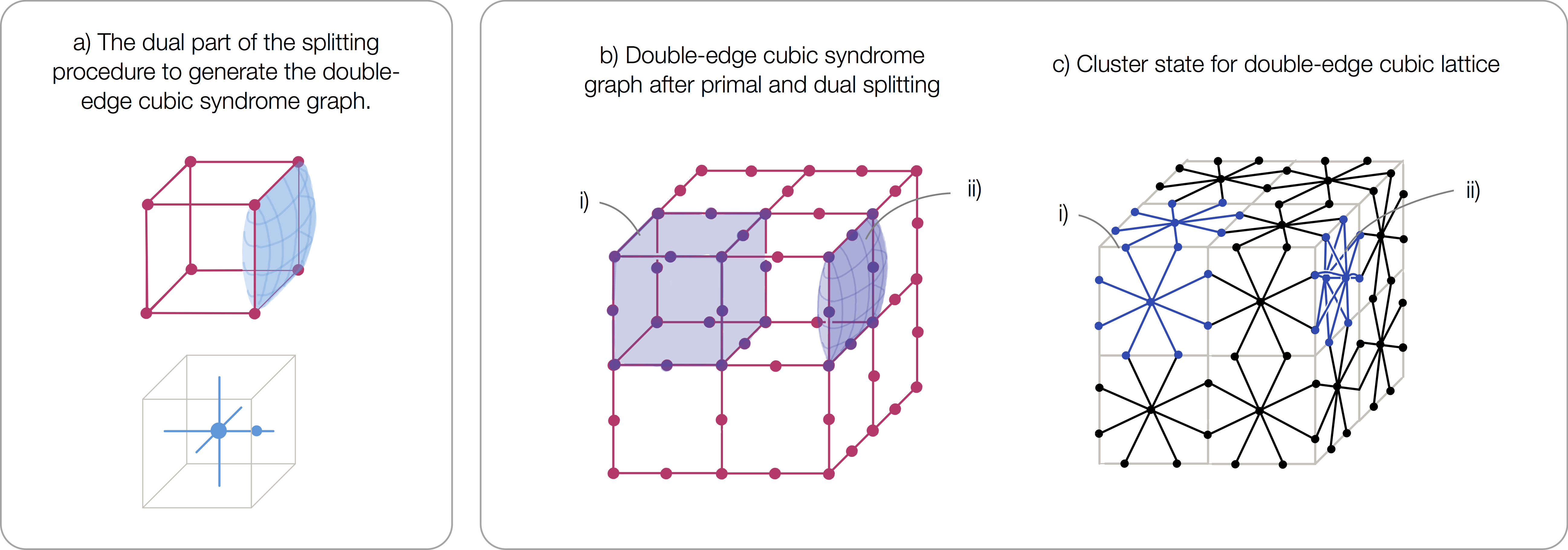}
\caption{\label{fig:double_edge_splitting} Splitting procedure to obtain a double-edge cubic syndrome graph. a) A single dual split along one edge is shown. In the dual lattice the original 6-valent vertex is transformed into two vertices, one is the unchanged 6-valent vertex, and the second is a 2-valent vertex that divides one of the original lattice edges into two. In the primal lattice this corresponds to adding a duplicate face to one side of the cubic cell. b) The resulting lattice structure after this splitting procedure is applied to affect every edge of the primal and dual lattice. The resulting structure is self-dual, but the cells are no longer all identical. There are two types of cells which are indicated in the figure. i) For every original cubic cell of the lattice there is a new cell which is made up of 6 octagonal faces. ii) For every face of the original cubic cell there is now a bubble-shaped cell, which has two octagonal faces. c) The cluster state defined by the new geometry of syndrome graph. Only external qubits and bonds are shown for clarity except for the one bubble-shaped cell labelled ii). Each qubit in the cluster state has 8 cluster state bonds. 
}
\end{figure}

The construction is shown in Figure~\ref{fig:double_edge_splitting}. Starting with a cubic syndrome graph, we can modify a single dual edge by adding a new vertex dividing it into two halves. In the primal lattice this corresponds to adding a duplicate face on one side of the cubic cell. By applying this transformation to every edge in the dual lattice, and every edge in the primal lattice we find the structure shown in Figure~\ref{fig:double_edge_splitting}b). Every face is octagonal, such that the cluster state is made up of qubits with 8 bonds, and the cell complex is again self-dual. Unlike our previous examples the cells are not all identical, there are two differently shaped types of cell as shown in the figure. A cell with 6 octagonal faces is associated with each cell of the original cubic lattice, and since each face has been doubled, there is also a bubble-shaped cell with just two octagonal faces associated with each face of the original lattice. We call this construction the {\em doubled-edge cubic FTCS}. An interesting observation about this state is that it can also be thought of as encoding each qubit of the original cubic FTCS in a 2-qubit parity code according to the "crazy-graph" construction described in~\cite{rudolph2017optimistic}.

The erasure threshold of this state can be inferred by considering how it relates to the original cubic lattice. For each edge of a cubic syndrome graph, there are now two edges, both of which must be erased in order to form a connected path. There is an effective erasure probability, $p_E^{\rm eff} = p_E^2$. The threshold against erasure is therefore $p_{\rm th} = \sqrt{0.249} \approx 0.5$. 

The same splitting procedure could be applied again by adding more vertices along each edge. If each of the primal and dual lattice are split in the same way the resultant cell complex is self-dual. For a state where the edges of the cubic lattice are divided by $n-1$ vertices into $n$ segments there is an erasure threshold of $p_{\rm th}= \sqrt[n]{0.249}$. The associated cluster state has qubits with $4n$ incident edges, so these higher order states are likely not realistic to construct. The fact that the threshold against erasure tends to $100\%$ demonstrates how the phenomenological error model is inappropriate for comparing these states in a meaningful way. 

\newpage

\section{Gate complementarity}

In the next section we will compute the threshold behavior of some examples from the class of fault tolerant states we have just introduced. Before we present these results we first briefly discuss the meaning of a threshold in a cluster state, as compared to the threshold of a code. The key conclusion is that if we can modify the resource state being created during the computation then there is no limit on the erasure threshold, while if we are only able to modify measurement patterns on a fixed resource state the erasure threshold is at most 50\%.

In MBQC the same resource state can be used to perform different computations by performing different sets of measurements. This is also true for fault-tolerant MBQC based on topological codes, where different sets of measurements can be chosen to perform different topological gates either based on code deformation~\cite{raussendorf2006fault} or transversal gates~\cite{bombin2015gauge}. In the absence of fault tolerance (roughly speaking) all the measurement outcomes are necessary, no erasures can happen. By contrast, in the topological scenario there exists an erasure error threshold: for erasure rates below the threshold the probability of logical errors approaches zero in the limit of large systems. The threshold depends on the measurement pattern, i.e. on the logical gate. Because the cluster state is consumed during measurement, only one logical gate can be performed, and we therefore have the following useful observation connecting error thresholds for different gates on the same resource state:

\vspace{.5\baselineskip}
\noindent {\bf Gate complementarity:} If a resource state can be used to implement two different gates with erasure thresholds $x$ and $y$, then $x+y\leq 1$.
\vspace{.5\baselineskip}

As we show in the next section, nothing prevents us from achieving erasure thresholds arbitrarily close to 100\%, but the price to pay is a lack of flexibility in the logical gates. The gate complementarity principle tells us that if we intend to perform different gates purely by changing measurement pattern on a fixed resource state, we cannot aim beyond 50\% erasure thresholds. For erasure thresholds beyond 50\% different gates must be enacted by also changing the resource state topology, which we note is perfectly compatible with certain architectures for MBQC such as photonic quantum computing.

It is worth pointing out that there is no contradiction between the possibility of having erasure thresholds beyond 50\% and the no-cloning theorem. A given measurement pattern on a subset of qubits $Q$ provides information on the correlations between the rest of qubits. Different parties with access to disjoint subsets $Q_i\subseteq Q$ might all be able to recover such classical information, with no contradiction as long as the correlations that they recover information about are the same.

\section{Threshold calculations}

\begin{figure}
\includegraphics[width=0.9\columnwidth]{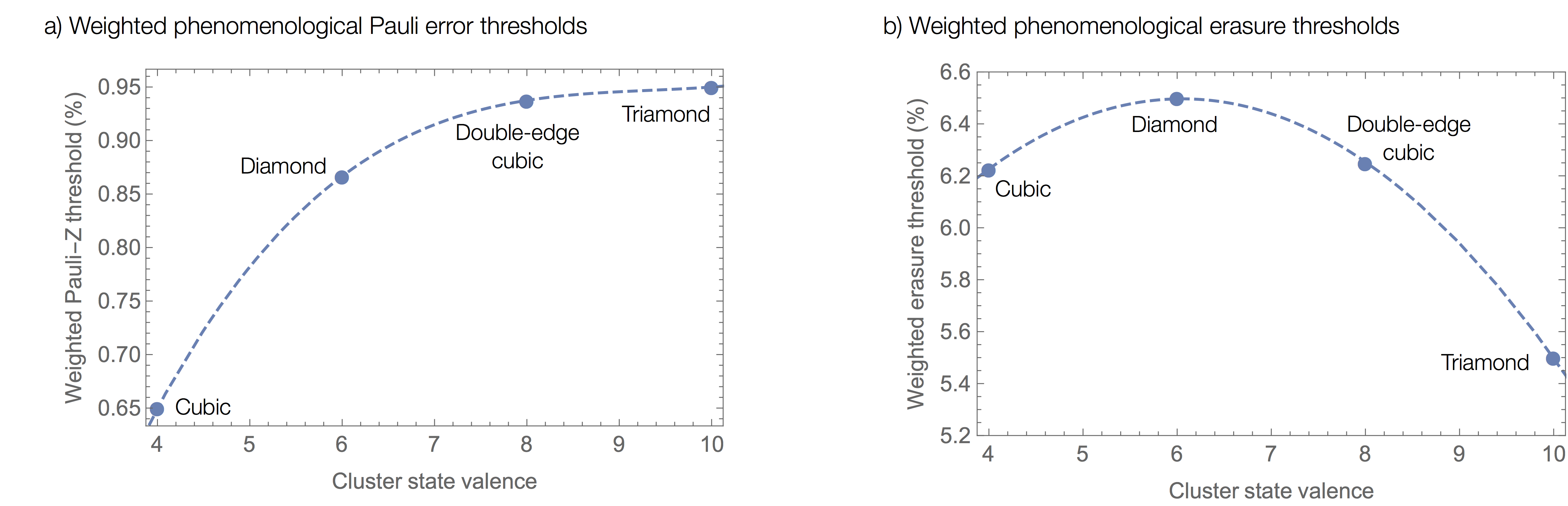}
\caption{\label{fig:weighted_thresholds} Weighted phenomenological thresholds, where each lattice site suffers a Pauli error with probability $p z$, where $p$ is a unit of error rate, and $z$ is the valence of the site in the cluster state.  a) Weighted Pauli thresholds as a function of the valence of the cluster state. b) Weighted erasure thresholds as a function of the valence of the cluster state.}
\end{figure}

We performed numerical simulations of each of the three FTCSs we introduced in the previous section under two phenomenological error models, one of erasure and the other of bit-flip Pauli errors. The erasure error model in the measurement based setting corresponds to perfect state preparation followed by measurements that return no outcome with probability $p_E$, whereas in the Pauli error model each measurement outcome is flipped with i.i.d. probability $p$. We use the union-find decoder~\cite{UF} because it can straightforwardly handle any geometry of syndrome graph, and any combination of Pauli error and erasure. The union-find decoder is optimal against erasure errors, but like all efficient decoders it does not perform with the optimal threshold against Pauli errors. For the cubic FTCS we know that the threshold is comparable to the threshold of the minimum weight perfect matching decoder (2.6\% vs. 2.9\%~\cite{wang2003confinement}), and is not far from the optimal threshold (3.3\%~\cite{ohno2004phase}) In general however is not known how the union-find decoder compares against other algorithms or optimal thresholds, so the Pauli error thresholds we find here are lower bounds. More details of the simulations are given in Appendix~\ref{app:numerics}, and the results are summarized in Table~\ref{table:summary} on page~\pageref{table:summary}. 

For the cubic FTCS our simulations reproduce the results of previous studies~\cite{barrett2010fault,UF}, with an erasure threshold of 24.9\% and a Pauli error threshold of 2.6\%. For the new cluster states we introduced we find a erasure thresholds of $39\%$ for the diamond FTCS, $50\%$ for the double-edge cubic FTCS, and $55\%$ for the triamond FTCS. These values are consistent with the known bond percolation thresholds~\cite{percolation3valent,bondpercolation}. We find a Pauli error threshold of $5.2\%$ for the diamond FTCS, $7.5\%$ for the double-edge cubic FTCS, and $9.5\%$ for the triamond FTCS. These results are consistent with the erasure threshold being a good proxy for the Pauli error threshold.

\subsection{Application to realistic quantum computing architectures}

We should now ask the important question of how these threshold numbers apply to a real quantum computing architecture. Phenomenological thresholds on their own do not tell us anything about tolerance to physical errors, as they only assess one side of the trade-off between the error tolerance of the perfect cluster state, and the difficulty of preparing that state. In Table~\ref{table:summary} we see that as the phenomenological thresholds go up, the valence of the cluster states also increases, meaning each qubit must go through more 2-qubit gates. We can make a fairer comparison by considering a {\em weighted phenomenological error model}, where error probability assigned to each qubit is weighted by the bond-degree of its site in the cluster state. Under this error model, we take a perfect cluster state and for each qubit, $q$, with cluster state valence $z_q$ the measurement result is flipped with probability $z_q p$, where $p$ is the error rate. This error model can give us a better idea of the behaviour under a gate based error model, and in the limiting case that all errors in the gates are phase errors the weighted phenomenological thresholds exactly reproduce the circuit model thresholds. The values for each FTCS are shown in Figure~\ref{fig:weighted_thresholds}. 

Without using a realistic error model derived from a the physical system and detailed simulation, no comparison of two codes can find out which would perform better in a real machine. We certainly cannot draw any concrete conclusions here on absolute performance for a real quantum computing architecture, and the answer will likely differ between devices. For example, the ratio of gate error rates and measurement error rates will have a big impact on the relative performance or different ratios of bit and phase errors in entangling operations could also change which FTCS has the higher threshold. However, we can conclude that the numbers under our comparison are close enough that it is worth analysing thee fault tolerant cluster state schemes under realistic noise.

For practical applications in quantum architectures there are considerations beyond the threshold, and it is important to emphasize that the measurement based approach we describe here can be implemented with matter based qubits~\cite{paler2014mapping,devitt2009architectural} where the states can prepared with 2-qubit gates, as well as in systems where there is no direct access to two-qubit gates, such as linear optics~\cite{kieling2007percolation,gimeno2015three,li2015resource}. In both cases, even though the 3-d fault tolerant cluster states cannot be interpreted as a single 2-d code, they can still be built with a 2-d and local physical architecture by preparing and measuring the cluster state layer by layer.

\section{Conclusions}

We have identified new fault-tolerant cluster states that cannot be constructed from the foliation of any 2-d surface code. While there is only one surface code geometry that is self-dual, by using the additional freedom of measurement based fault tolerance in the third dimension we have found multiple new self-dual fault tolerant cluster states and in doing so we can reduce the valence of both the primal and dual syndrome graph. As a way of thinking about fault-tolerance the measurement based approach brings greater flexibility, giving a clear picture of a fault tolerant map as opposed to a static error correcting code, and a more natural insight into the structure of the syndrome information. Our numerical results reinforce that the bond-degree of a syndrome graph is the primary indicator of tolerance to erasure errors, and secondly that there is indeed a strong relationship between erasure thresholds and Pauli error thresholds. With the example of the triamond FTCS we have found we can indeed construct a self-dual example in three-dimensions which achieves the minimum degree for a syndrome graph. 

It is obviously a primary goal to find codes with higher tolerance to errors, and indeed in all the examples we introduce here the thresholds for erasure and Pauli error under a phenomenological error model are indeed significantly higher than those of any foliated surface code (Table~\ref{table:summary}), and even exceed 50\%. Even when the preparation of the cluster state is considered, our results indicate promising thresholds that merit further comparison under the specific constraints and noise models of a realistic quantum architecture to determine under which scenarios these geometric modifications can give a real performance advantage. 

In making such comparisons it is also worth noting that the threshold values we have found here can likely be improved upon by using different decoding algorithms. Here we used the basic union find decoder, which was a simple approach given its adaptability to any geometry. For the cubic FTCS we know that the UF decoder is competitive with the minimum weight perfect-matching decoder, but for the other thresholds it is not clear how their performance will differ. Beyond practical thresholds using efficient decoders, it would also be interesting to determine their optimal performance using the mappings to statistical mechanics models that have been used to study other topological codes~\cite{ohno2004phase,andrist2011tricolored,kubica2017three}.

The method of splitting that we introduced and used to construct these new cluster states is a general purpose tool that can be used to explore many more geometries of surface code cluster states. Furthermore similar ideas could potentially be applied to other classes of fault tolerant cluster states. If we were to highlight one key takeaway from this study, it would be that geometry still has a role to play in the exploration of better schemes for fault tolerance.

\medskip
{\em Acknowledgements -- The authors would like to thank the  whole  PsiQuantum  fault  tolerance  team: Chris Dawson, Fernando  Pastawski,  Nicolas  Breuckmann, Kiran  Mathew,  Andrew Doherty, Jordan Sullivan, and Mihir Pant, as well as Eric Johnson, Sara Bartolucci and Pete Shadbolt for their  constant  support and  encouragement. We would also like to thank Tom Stace and Benjamin Brown for many valuable discussions. In particular we thank  Mercedes Gimeno-Segovia and Terry Rudolph for the introduction to the loss-tolerant code that inspired the doubled-edge cubic FTCS and Chris Dawson for the implementation of the decoder.}

\bibliographystyle{apsrev}
\bibliography{vcodes}

\newpage
\clearpage

\begin{appendix}

\onecolumngrid

\section{Threshold Numerics}
\label{app:numerics}

We performed numerical simulations of error recovery on the three FTCSs described in the main text: the diamond FTCS, the triamond FTCS and the doubled-edge cubic FTCS. For each state we simulated the performance under two error models. Firstly, under a phenomenological error model where each qubit has an i.i.d probability $p$ of suffering a Pauli-Z error (bit-flip) prior to measurement. An equivalent model is that each single qubit measurement returns the incorrect outcome with probability $p$. In the second error model the Pauli-Z error rate $p=0$, and there is an i.i.d probability $p_E$ of an erasure error on any given qubit. Equivalently, this can be stated as perfect state preparation, followed by measurement which is either perfect, or returns an invalid outcome with probability $p_E$. 

We consider a syndrome graph with periodic boundary conditions in all directions. For each FTCS we simulate four lattice sizes, $L$, where the full lattice is made of of $L$ unit cells in each of the 3 dimensions, where we take as the unit cells those based on the  cubic structure as shown in Figures~\ref{fig:diamond},\ref{fig:triamond} and \ref{fig:double_edge_splitting}.

\begin{figure*}[h]
	\includegraphics[width=\columnwidth]{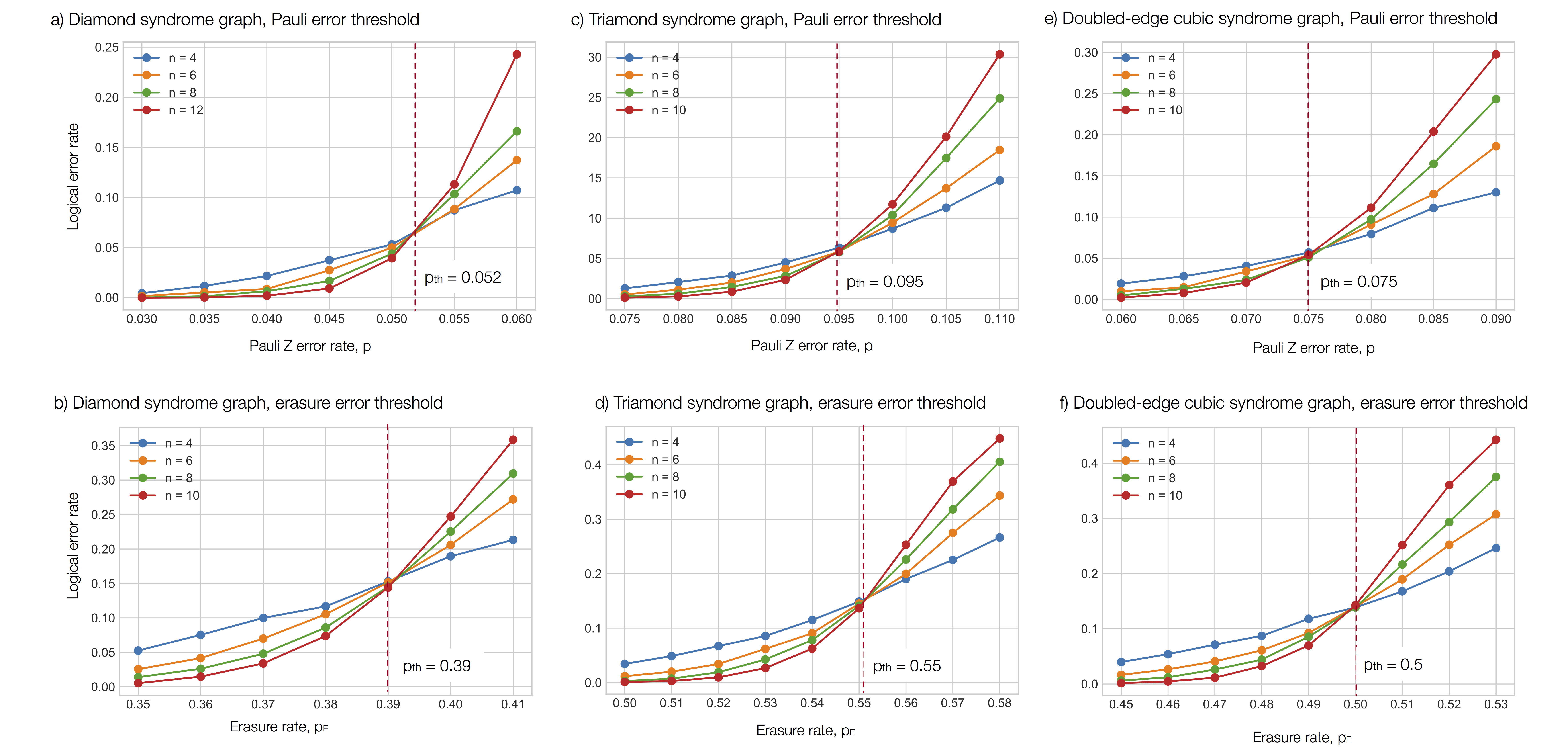}
	\caption{\label{fig:thresholds} Threshold results. Montecarlo simulations of decoding under Pauli error and erasure were performed on a each of the three syndrome graphs described in the main text. For each cluster state threshold calculations are shown under two error models. In the first Pauli-Z errors are applied with i.i.d probability $p$, and there are. no erasures, $p_E=0$. In the second, there are no Pauli errors, and erasures occur with probability $p_E$. Each data point shown is the result of at least $10^4$ trials. Decoding was performed using the union-find decoder~\cite{UF} with uniform cluster growth. a) Pauli error threshold for the diamond syndrome graph. We see a threshold crossing at $p_Z = 5.2\%$. b) Erasure threshold for the diamond syndrome graph. We see a threshold crossing at $p_E=0.39$, which is consistent with known percolation results in the literature~\cite{bondpercolation}. c) Pauli error threshold for the triamond syndrome graph. We see a threshold crossing at $p_Z = 9.5\%$. d) Erasure threshold for the triamond syndrome graph. We see a threshold crossing at $p_E=0.55$, which is consistent with known percolation results in the literature~\cite{bondpercolation}. e) Pauli error threshold for the doubled-edge syndrome graph. The threshold crossing is seen at $p_Z=7.5\%$. f) Erasure threshold for the doubled-edge syndrome graph. The threshold crossing is seen at $p=50\%$. This is consistent with an analytic comparison with the cubic syndrome graph in which we square the erasure probability along any given edge.
	}
\end{figure*}

Each numerical simulation proceeds by generating a random error configuration on the syndrome graph, and applying the union-find decoding algorithm described in Ref.~\cite{UF} to find a correction. After this correction is applied we measure a single logical operator and determine whether the decoding attempt was a success or failure. We repeat this process at least $10^4$ times for each combination of error probability $p$ or $p_E$ and lattice size, $L$. For reference we run the decoding algorithm on the cubic syndrome graph, which corresponds to the cubic FTCS. We find a threshold against erasure of $p_{\rm th}^E=0.25$, and a threshold against Pauli-Z error of $p_{\rm th}=0.026$. These numerics are consistent with previous reported results
\cite{barrett2010fault}. The results of Montecarlo simulations of the diamond, triamond and doubled-edge cubic FTCSs are shown in Figure~\ref{fig:thresholds}.

\end{appendix}

\end{document}